\documentclass{PoS}
\usepackage{graphics}

\title{The spectrum of static-light baryons in twisted mass lattice QCD}

\ShortTitle{The spectrum of static-light baryons in twisted mass lattice QCD}

\author{Marc Wagner, \speaker{Christian Wiese}\\
        Humboldt-Universit\"at zu Berlin, Institut f\"ur Physik, Newtonstra{\ss}e 15, D-12489 Berlin, Germany\\
        E-mail: \email{mcwagner@physik.hu-berlin.de}\\
        E-mail: \email{wiese@physik.hu-berlin.de}}

\abstract{
We compute the static-light baryon spectrum with $N_f = 2$ flavors of sea quarks using Wilson twisted mass lattice QCD. As light valence quarks we consider quarks, which have the same mass as the sea quarks with corresponding pion masses in the range $340 \, \textrm{MeV} \ltapprox m_\mathrm{PS} \ltapprox 525 \, \textrm{MeV}$, as well as partially quenched quarks, which have the mass of the physical $s$ quark. We extract masses of states with isospin $I = 0, 1/2, 1$, with strangeness $S = 0, -1, -2$, with angular momentum of the light degrees of freedom $j = 0, 1$ and with parity $\mathcal{P} = +, -$. We present a preliminary extrapolation in the light $u/d$ and an interpolation in the heavy $b$ quark mass to the physical point and compare with available experimental results.
}

\FullConference{The XXVIII International Symposium on Lattice Field Theory, Lattice2010\\
		June 14-19, 2010\\
		Villasimius, Italy}

\newcommand{\ltapprox}{\raisebox{-0.5ex}{$\,\stackrel{<}{\scriptstyle\sim}\,$}}

\begin{document}


\section{Introduction}

A systematic way to study bottom baryons from first principles is lattice QCD. Since $a m_b > 1$ for currently available lattice spacings for large volume simulations, one needs to use for the $b$ quark a formalism such as Heavy Quark Effective Theory (HQET). In this paper we consider the leading order of HQET, which is the static limit. The spectrum of static-light baryons has been studied by lattice methods in the quenched approximation \cite{Ewing:1995ih,Michael:1998sg,Detmold:2007wk} and recently also with dynamical sea quarks \cite{Burch:2008qx,Detmold:2008ww,Lin:2010wb}. Here we use $N_f=2$ flavors of dynamical Wilson twisted mass quarks. Our intention is to explore the static-light baryon spectrum as fully as possible. Besides the experimentally known states $\Lambda_b^0$, $\Sigma_b / \Sigma_b^\ast$, $\Xi_b^-$ and $\Omega_b^-$ we also predict a number of not yet measured states, mainly of negative parity.


\section{Simulation setup}

We use $24^3 \times 48$ gauge field configurations generated by the European Twisted Mass Collaboration (ETMC). The fermion action is $N_f = 2$ Wilson twisted mass \cite{Frezzotti:2000nk,Frezzotti:2003ni} tuned to maximal twist, where static-light mass differences are automatically $\mathcal{O}(a)$ improved \cite{Jansen:2008si}. The gauge action is tree-level Symanzik improved \cite{Weisz:1982zw} with $\beta = 3.9$ corresponding to a lattice spacing $a = 0.079(3) \, \textrm{fm}$ \cite{Baron:2009wt}. At the moment we have considered three different values of the twisted mass $\mu_\mathrm{q}$ with corresponding pion masses listed in Table~\ref{TAB001}. For details regarding these gauge field configurations we refer to \cite{Boucaud:2007uk,Boucaud:2008xu}.

In the valence sector we use light quarks, which have the same mass as the sea quarks (corresponding to $u/d$ quarks), as well as partially quenched quarks with a mass $\mu_\mathrm{q,val} = 0.0220$, which is around the mass of the physical $s$ quark \cite{Blossier:2007vv,Blossier:2009bx}.





\begin{table}[htb]
\begin{center}
\begin{tabular}{|c|c|c|}
\hline
 & & \vspace{-0.40cm} \\
$\mu_\mathrm{q}$ & $m_\mathrm{PS}$ in MeV & number of gauge field configurations \\
 & & \vspace{-0.40cm} \\
\hline
 & & \vspace{-0.40cm} \\
$0.0040$ & $340(13)$ & $200$ \\
$0.0064$ & $423(16)$ & $\phantom{0}40$ \\
$0.0100$ & $525(20)$ & $\phantom{0}30$\vspace{-0.40cm} \\
 & & \\
\hline
\end{tabular}
\caption{\label{TAB001}twisted masses $\mu_\mathrm{q}$, corresponding pion masses $m_\mathrm{PS}$ and number of gauge field configurations considered.}
\end{center}
\end{table}


\section{Static-light baryon creation operators}

To create static-light baryons, we use operators
\begin{eqnarray}
\label{EQN002} \mathcal{O}_\Gamma^{\scriptsize \textrm{twisted}} \ \ = \ \ \epsilon^{a b c} Q^a \Big((\chi^{b,(1)})^T \mathcal{C} \Gamma^{\scriptsize \textrm{twisted}} \chi^{c,(2)}\Big) ,
\end{eqnarray}
where $Q$ is a static quark operator and $\chi^{(n)}$ are light quark operators in the so-called twisted basis. The upper indices $a$, $b$ and $c$ are color indices, $\mathcal{C} = \gamma_0 \gamma_2$ is the charge conjugation matrix and $\Gamma^{\scriptsize \textrm{twisted}}$ is an appropriately chosen combination of $\gamma$ matrices.

When discussing quantum numbers of static-light baryons, it is more convenient to first transform the operators (\ref{EQN002}) into the physical basis, in which they have the same structure, i.e.\
\begin{eqnarray}
\mathcal{O}_\Gamma^{\scriptsize \textrm{physical}} \ \ = \ \ \epsilon^{a b c} Q^a \Big((\psi^{b,(1)})^T \mathcal{C} \Gamma^{\scriptsize \textrm{physical}} \psi^{c,(2)}\Big) .
\end{eqnarray}
In the continuum the relation between the physical and the twisted basis is given by the twist rotation $\psi = \exp(i \gamma_5 \tau_3 \omega / 2) \chi$, where the twist angle $\omega = \pi / 2$ at maximal twist. At finite lattice spacing, however, issues are more complicated: the twist rotation only holds for renormalized operators and the QCD symmetries isospin and parity are explicitely broken by $\mathcal{O}(a)$. Nevertheless, it is possible to unambiguously interpret states obtained from correlation functions of twisted basis operators in terms of QCD quantum numbers. The method has successfully been applied in the context of static-light mesons \cite{Blossier:2009vy} and is explained in detail for kaons and $D$ mesons in \cite{Baron:2010th}. For details regarding its application to static-light baryons we refer to an upcoming publication \cite{WW2010}.

Since there are no interactions involving the static quark spin, it is appropriate to label static-light baryons by angular momentum $j$ and parity $\mathcal{P}$ of the light degrees of freedom, which are determined by $\Gamma^{\scriptsize \textrm{physical}}$. Consequently, $j = 1$ states correspond to degenerate pairs of states with total angular momentum $J = 1/2$ and $J = 3/2$ respectively.

To access all possible isospin and strangeness combinations, we consider light quark flavors $\psi^{(1)} \psi^{(2)} = ud - du$ (corresponding to $I = 0$, $S = 0$), $\psi^{(1)} \psi^{(2)} \in \{ uu \, , \, dd \, , \, ud + du \}$ (corresponding to $I = 1$, $S = 0$), $\psi^{(1)} \psi^{(2)} \in \{ us \, , \, ds \}$ (corresponding to $I = 1/2$, $S = -1$) and $\psi^{(1)} \psi^{(2)} = ss$ (corresponding to $I = 0$, $S = -2$).

The operators we have considered and their associated quantum numbers are collected in Table~\ref{TAB002}.

\begin{table}[htb]
\begin{center}
\begin{tabular}{|c||c|c||c|c|c||c|c|c||c|c|c|}
\hline
 & & & & & & & & & & & \vspace{-0.40cm} \\
$\Gamma^{\scriptsize \textrm{physical}}$ & $j^\mathcal{P}$ & $J$ & $I$ & $S$ & name & $I$ & $S$ & name & $I$ & $S$ & name \\
 & & & & & & & & & & & \vspace{-0.40cm} \\
\hline
 & & & & & & & & & & & \vspace{-0.40cm} \\
$\gamma_5$ & $0^+$ & $1/2$ & $0$ & $0$ & $\Lambda_b^0$ & $1/2$ & $-1$ & $\Xi_b^-$ & X & X & X \\
$\gamma_0 \gamma_5 $ & $0^+$ & $1/2$ & $0$ & $0$ & $\Lambda_b^0$ & $1/2$ & $-1$ & $\Xi_b^-$ & X & X & X \\
$1$ & $0^-$ & $1/2$ & $0$ & $0$ & $-$ & $1/2$ & $-1$ & $-$ & X & X & X \\
$\gamma_0$ & $0^-$ & $1/2$ & $1$ & $0$ & $-$ & $1/2$ & $-1$ & $-$ & $0$ & $-2$ & $-$ \\
 & & & & & & & & & & & \vspace{-0.40cm} \\
\hline
 & & & & & & & & & & & \vspace{-0.40cm} \\
$\gamma_j$ & $1^+$ & $1/2$, $3/2$ & $1$ & $0$ & $\Sigma_b$, $\Sigma_b^\ast$ & $1/2$ & $-1$ & $-$ & $0$ & $-2$ & $\Omega_b^-$ \\
$\gamma_0 \gamma_j$ & $1^+$ & $1/2$, $3/2$ & $1$ & $0$ & $\Sigma_b$, $\Sigma_b^\ast$ & $1/2$ & $-1$ & $-$ & $0$ & $-2$ & $\Omega_b^-$ \\
$\gamma_j \gamma_5$ & $1^-$ & $1/2$, $3/2$ & $0$ & $0$ & $-$ & $1/2$ & $-1$ & $-$ & X & X & X \\
$\gamma_0 \gamma_j \gamma_5$ & $1^-$ & $1/2$, $3/2$ & $1$ & $0$ & $-$ & $1/2$ & $-1$ & $-$ & $0$ & $-2$ & $-$\vspace{-0.40cm} \\
 & & & & & & & & & & & \\
\hline
\end{tabular}
\caption{\label{TAB002}static-light baryon creation operators and their quantum numbers ($j^\mathcal{P}$: angular momentum and parity of the light degrees of freedom; $J$: total angular momentum; $I$: isospin; $S$: strangeness); operators marked with ``X'' are identically zero, i.e.\ do not exist.}
\end{center}
\end{table}


\section{Numerical results}

We compute correlation matrices
\begin{eqnarray}
\label{EQN001} C_{\Gamma_j,\Gamma_k}(t) \ \ = \ \ \langle \Omega | \mathcal{O}_{\Gamma_j}^{\scriptsize \textrm{twisted}}(t) \Big(\mathcal{O}_{\Gamma_k}^{\scriptsize \textrm{twisted}}(0)\Big)^\dagger | \Omega \rangle .
\end{eqnarray}
We use several techniques to improve the signal quality, including operator optimization by means of APE and Gaussian smearing, the HYP2 static action and stochastic propagators combined with timeslice dilution. These techniques are very similar to those used in a recent study of the static-light meson spectrum \cite{Jansen:2008si,:2010iv} and will be explained in detail in \cite{WW2010}.

\begin{figure}[htb]
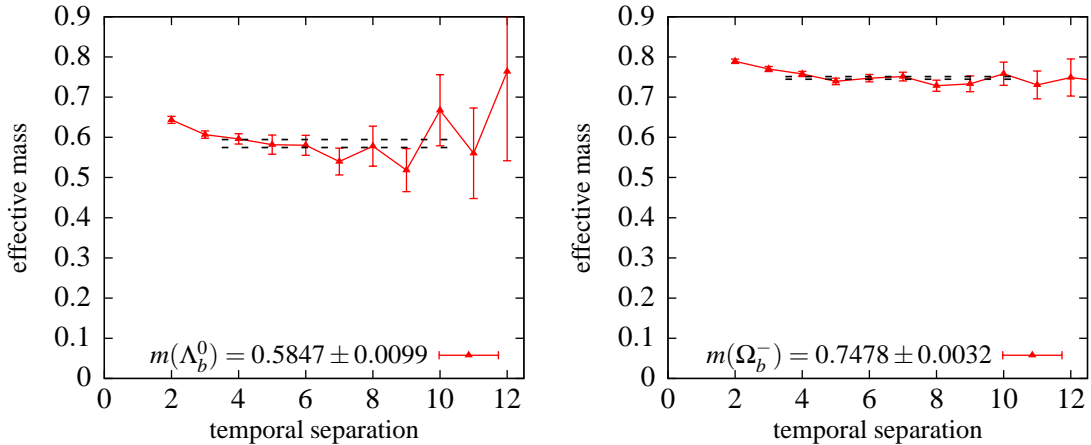

\begin{center}
\input{ud_scalar1.tex}\input{omega.tex}
\caption{\label{FIG001}effective masses and corresponding mass fits; left: $\Lambda_b^0$ at $\mu_\mathrm{q} = 0.0040$ from a $3 \times 3$ correlation matrix; right: $\Omega_b^-$ at $\mu_\mathrm{q} = 0.0040$ from a $3 \times 3$ correlation matrix.}
\end{center}
\end{figure}

From the correlation matrices (\ref{EQN001}) we extract effective masses by solving a generalized eigenvalue problem (cf.\ e.g.\ \cite{Blossier:2009kd} and references therein). Mass values are obtained by fitting constants to effective mass plateaus at sufficiently large temporal separations. To exemplify the quality of our results, we show effective masses and corresponding mass fits for the $\Lambda_b^0$ and the $\Omega_b^-$ baryon in Figure~\ref{FIG001}.

Static-light baryon masses diverge in the continuum limit, because of the infinite self energy of the static quark. Therefore, we consider mass differences to another static-light system, the lightest static-light meson (quantum numbers $j^\mathcal{P} = (1/2)^-$, corresponding to the $B$ and the $B^\ast$ meson, which are degenerate in the static limit). In such mass differences the self energies of the static quarks exactly cancel.

\begin{figure}[htb]
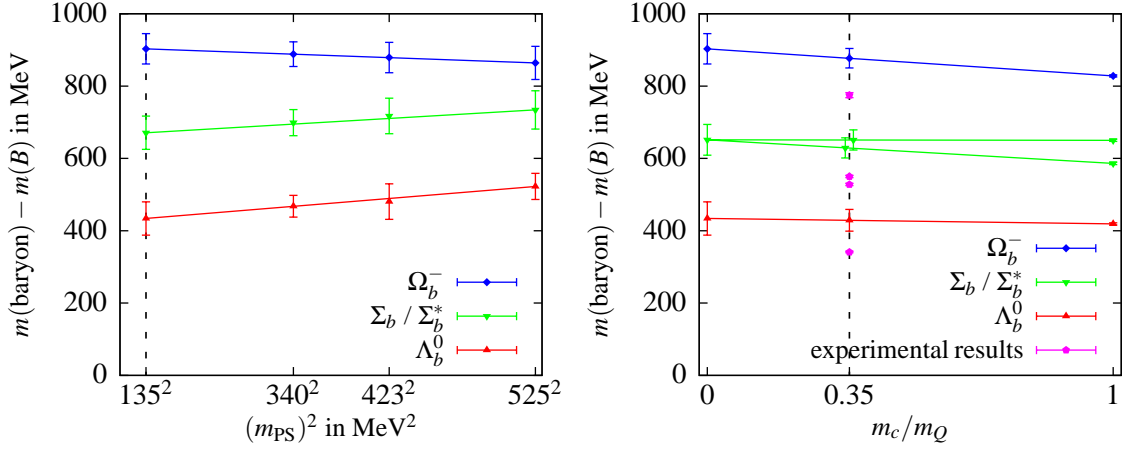

\begin{center}
\input{extra.tex}\input{inter.tex}
\caption{\label{FIG002}left: linear extrapolation of static-light mass differences in $(m_\mathrm{PS})^2$ to the physical $u/d$ quark mass corresponding to $m_\mathrm{PS} = 135 \, \textrm{MeV}$; right: linear interpolation between lattice static-light and corresponding experimental charm-light mass differences in $m_c/m_Q$ to the physical $b$ quark mass corresponding to $m_c / m_b = 0.35$.}
\end{center}
\end{figure}

We use our results at three different light quark masses (cf.\ Table~\ref{TAB001}) by extrapolating mass differences linearly in $(m_\mathrm{PS})^2$ to the physical $u/d$ quark mass ($m_\mathrm{PS} = 135 \, \textrm{MeV}$). Figure~\ref{FIG002} (left) shows this extrapolation for the $\Lambda_b^0$, the $\Sigma_b/\Sigma_b^\ast$ and the $\Omega_b^-$.

Our results for the static-light baryon spectrum are collected in Table~\ref{TAB003}. In Wilson twisted mass lattice QCD isospin is explicitely broken by $\mathcal{O}(a)$. Therefore, states with $I \neq 0$ and different $I_z$ will have slightly different masses differing by $\mathcal{O}(a^2)$. Because of this, in Table~\ref{TAB003} two values are listed for such states. Note, however, that within statistical errors there is no difference between any two such values, which is in agreement with the expectation that isospin breaking effects are $\mathcal{O}(a^2)$ and, hence, should be rather small.

\begin{table}[htb]
\begin{center}
\begin{tabular}{|c|c|c|c|c|}
\hline
 & & & & \vspace{-0.40cm} \\
$j^\mathcal{P}$ & $I$ & $S$ & name & $m(\textrm{baryon}) - m(B)$ in $\textrm{MeV}$ \\
 & & & & \vspace{-0.40cm} \\
\hline
 & & & & \vspace{-0.40cm} \\
%
$0^+$ & $0$ & $0$ & $\Lambda_b^0$ & $434(46)$ \\
%
$1^+$ & $1$ & $0$ & $\Sigma_b / \Sigma_b^\ast$ & $671(46) / 632(39)$ \\
%
$0^-$ & $0$ & $0$ & $-$ & $1389(113)$ \\
%
$1^-$ & $1$ & $0$ & $-$ & $1008(92) / 1014(79)$ \\
 & & & & \vspace{-0.40cm} \\
\hline
 & & & & \vspace{-0.40cm} \\
%
$0^+$ & $1/2$ & $-1$ & $\Xi_b^-$ & $630(41) / 677(36)$ \\
%
$1^+$ & $1/2$ & $-1$ & $-$ & $789(45) / 798(49)$ \\
%
$0^-$ & $1/2$ & $-1$ & $-$ & $1200(90) / 1262(77)$ \\
%
$1^-$ & $1/2$ & $-1$ & $-$ & $1233(58) / 1285(69)$ \\
 & & & & \vspace{-0.40cm} \\
\hline
 & & & & \vspace{-0.40cm} \\
%
$1^+$ & $0$ & $-2$ & $\Omega_b^-$ & $903(42)$ \\
%
$1^-$ & $0$ & $-2$ & $-$ & $1315(79)$\vspace{-0.40cm} \\
 & & & & \\
\hline
\end{tabular}
\caption{\label{TAB003}mass differences of static-light baryons and the lightest static-light meson extrapolated to the physical $u/d$ quark mass.}
\end{center}
\end{table}


\section{Interpolation in the heavy quark mass and comparison to experimental data}

To make predictions regarding the spectrum of $b$ baryons and to perform a meaningful comparison with experimental data, we interpolate linearly $m_c / m_Q$, where $m_Q$ is the heavy quark mass, to the physical point $m_c / m_b \approx m(D) / m(B) = 0.35$. We do this by using our static-light lattice results (cf.\ Table~\ref{TAB003}) corresponding to $m_c / m_Q = 0$ and rather precise experimental data for corresponding charm-light systems corresponding to $m_c / m_Q = 1$. In Figure~\ref{FIG002} (right) we show the interpolation for $\Lambda_b^0$, $\Sigma_b/\Sigma_b^\ast$ and $\Omega_b^-$.

In Table~\ref{TAB004} we compare results of these interpolations with experimentally available data for $b$ baryons. Our lattice results are around $15 \%$ larger than the corresponding experimental results, a tendency, which has already been observed in our recent study of the static-light meson spectrum \cite{Jansen:2008si,:2010iv}. In view of this discrepancy it is interesting to compare with results obtained by other lattice groups. When comparing mass differences in units of $r_0$, i.e.\ dimensionless quantities, e.g.\ with \cite{Burch:2008qx}, we find agreement within statistical errors. Note, however, that the scale setting is rather different: $r_0 = 0.49 \, \textrm{fm}$ in \cite{Burch:2008qx}, while the corresponding ETMC value is $r_0 = 0.42 \, \textrm{fm}$.

\begin{table}[htb]
\begin{center}
\begin{tabular}{|c|c|c|}
\hline
 & & \vspace{-0.40cm} \\
name & $m(\textrm{baryon}) - m(B)$ in $\textrm{MeV}$ (lattice) & $m(\textrm{baryon}) - m(B)$ in $\textrm{MeV}$ (experiment) \\
 & & \vspace{-0.40cm} \\
\hline
 & & \vspace{-0.40cm} \\
$\Lambda_b^0$ & $429(30)$ & $341(2) \quad$ (from \cite{PDG}) \\
$\Sigma_b$ & $629(28)$ & $528(3) \quad$ (from \cite{PDG}) \\
$\Sigma_b^\ast$ & $651(28)$ & $550(3) \quad$ (from \cite{PDG}) \\
$\Xi_b^-$ & $635(25)$ & $513(3) \quad$ (from \cite{PDG}) \\
$\Omega_b^-$ & $877(27)$ & $775(7) \quad$ (from \cite{Aaltonen:2009ny})\vspace{-0.40cm} \\
 & & \\
\hline
\end{tabular}
\caption{\label{TAB004}lattice predictions versus experimental results for mass differences between various $b$ baryons and the $B$ meson.}
\end{center}
\end{table}

For the negative parity states listed in Table~\ref{TAB003} experimental results for $b$ baryons do not exist. There are also no lattice predictions for these states we are aware of. However, one can compare with phenomenological models, e.g.\ with \cite{Ebert:2007nw}, for which one finds qualitative agreement.


\section{Conclusions}

We have studied the static-light baryon spectrum by means of $N_f = 2$ Wilson twisted mass lattice QCD. We have considered three different values of the dynamical quark mass corresponding to $340 \, \textrm{MeV} \ltapprox m_\mathrm{PS} \ltapprox 525 \, \textrm{MeV}$. We have performed an extrapolation in the light quark mass to the physical $u/d$ mass as well as an interpolation in the heavy quark mass to the physical $b$ mass. Our results agree within around $15 \%$ with currently available experimental results for $b$ baryons.

Future plans regarding this project include increasing the statistical accuracy of our correlation matrices to a level, where also excited states can reliably be extracted. Moreover, we intend to investigate the continuum limit, which amounts to considering other finer values of the lattice spacing. Finally we plan to perform similar computations on $N_f = 2+1+1$ flavor gauge field configurations currently produced by ETMC \cite{Baron:2010bv}.


\begin{acknowledgments}

We acknowledge useful discussions with Jaume Carbonell, William Detmold, Vladimir Galkin, Karl Jansen, Andreas Kronfeld, Chris Michael and Michael M\"uller-Preussker. This work has been supported in part by the DFG Sonderforschungsbereich TR9 Computergest\"utzte The\-o\-re\-tische Teilchenphysik.

\end{acknowledgments}



\end{document}